\documentclass[12pt,epsf]{article}
\usepackage{graphicx}
\usepackage{epsfig,amsmath,amssymb,verbatim,mathrsfs,subfigure,feynmf}
\begin{document}
%%%%%%%%%%%%%%%%%%%%%%%%%%%%%%%%%%%%%%%%%%%

\def\a{\alpha}
\def\b{\beta}
\def\c{\varepsilon}
\def\d{\delta}
\def\e{\epsilon}
\def\f{\phi}
\def\g{\gamma}
\def\h{\theta}
\def\k{\kappa}
\def\l{\lambda}
\def\m{\mu}
\def\n{\nu}
\def\p{\psi}
\def\q{\partial}
\def\r{\rho}
\def\s{\sigma}
\def\t{\tau}
\def\u{\upsilon}
\def\v{\varphi}
\def\w{\omega}
\def\x{\xi}
\def\y{\eta}
\def\z{\zeta}
\def\D{\Delta}
\def\G{\Gamma}
\def\H{\Theta}
\def\L{\Lambda}
\def\F{\Phi}
\def\P{\Psi}
\def\S{\Sigma}

\def\o{\over}
\def\beq{\begin{eqnarray}}
\def\eeq{\end{eqnarray}}
\newcommand{\gsim}{ \mathop{}_{\textstyle \sim}^{\textstyle >} }
\newcommand{\lsim}{ \mathop{}_{\textstyle \sim}^{\textstyle <} }
\newcommand{\vev}[1]{ \left\langle {#1} \right\rangle }
\newcommand{\bra}[1]{ \langle {#1} | }
\newcommand{\ket}[1]{ | {#1} \rangle }
\newcommand{\EV}{ {\rm eV} }
\newcommand{\KEV}{ {\rm keV} }
\newcommand{\MEV}{ {\rm MeV} }
\newcommand{\GEV}{ {\rm GeV} }
\newcommand{\TEV}{ {\rm TeV} }
\def\diag{\mathop{\rm diag}\nolimits}
\def\Spin{\mathop{\rm Spin}}
\def\SO{\mathop{\rm SO}}
\def\O{\mathop{\rm O}}
\def\SU{\mathop{\rm SU}}
\def\U{\mathop{\rm U}}
\def\Sp{\mathop{\rm Sp}}
\def\SL{\mathop{\rm SL}}
\def\tr{\mathop{\rm tr}}

\def\IJMP{Int.~J.~Mod.~Phys. }
\def\MPL{Mod.~Phys.~Lett. }
\def\NP{Nucl.~Phys. }
\def\PL{Phys.~Lett. }
\def\PR{Phys.~Rev. }
\def\PRL{Phys.~Rev.~Lett. }
\def\PTP{Prog.~Theor.~Phys. }
\def\ZP{Z.~Phys. }
\newcommand{\draftnote}[1]{\textbf{#1}}

%%%%%%%%%%%%%%%%%%%%%%%%%%%%%%%%%%%%%%%%%%%%%%%%%%%%%%%%%%%%%%%%%%%%

\baselineskip 0.7cm

\begin{titlepage}

\begin{flushright}
IPMU11-0067
\end{flushright}

\vskip 1.35cm
\begin{center}
{\large \bf  Hermitian Flavor Violation}
\vskip 1.2cm
%Jason Evans, Brian Feldstein, William Klemm, Hitoshi Murayama and  Tsutomu T. Yanagida
%\vskip 0.4cm

%{\it  Department of Physics,USA \\
 %   IPMU,  Japan \\}

Jason L. Evans$^{a}$, Brian Feldstein$^{a}$,  William Klemm$^{a,b,c}$,\\ Hitoshi Murayama$^{a,b,c}$ and Tsutomu T. Yanagida$^{a,d}$
\vskip 0.4cm

${}^{a}${\it  Institute for the Physics and Mathematics of the Universe, \\
University of Tokyo, Kashiwa, 277-8568, Japan}\\
${}^{b}${\it Department of Physics, University of California, Berkeley, CA 94720, USA}\\
${}^{c}${\it Theoretical Physics Group, Lawrence Berkeley National Laboratory,\\ Berkeley, CA 94720, USA}\\
${}^{d}${\it Department of Physics, University of Tokyo, Tokyo 113-0033, Japan}

\vskip 1.5cm

\abstract{The fundamental constraint on two Higgs doublet models comes from
the requirement of sufficiently suppressing flavor-changing neutral currents.   There are various standard
approaches for dealing with this problem, but they all tend to share a common feature;  all of the Higgs doublets
couple very weakly to the first generation quarks.  Here we consider a simple two Higgs doublet model
which is able to have large couplings to the first generation, while also being safe from flavor constraints.  We assume
only that there is an $SU(3)_f$ flavor symmetry which is respected by the couplings of one of the Higgs doublets, and which is broken by Hermitian Yukawa couplings of the second doublet.  As a result of the large permitted couplings to the first generation quarks, this scenario may be used to address the excess in $W+$dijet events recently observed by CDF at the Tevatron. Moreover, Hermitian Yukawa coupling matrices arise naturally in a broad class of
solutions to the strong CP problem, providing a compelling context for the model.

}
%\abstract{ Natural Flavor Conservation in the Higgs sector is usually
 %believed to allow only two types of twou-doublet Higgs models,
 % Type-I and II.  We point that there is yet another possibility with
 % $O(1)$ Yukawa couplings to a new Higgs which naturally satisfies all
  %flavor constraints.  It is loosely based on the Nelson--Barr
  %scenario of the strong CP problem.}
\end{center}
\end{titlepage}

\setcounter{page}{2}

\section{Introduction}

Models with multiple Higgs bosons provide one of the simplest
possibilities for physics beyond the standard model.  Indeed, two
Higgs doublet models in particular have received a great deal of
attention, and have arisen in a wide variety of contexts, including
supersymmetry or extra dimensions, as well as axion models.  More
generally, given our current lack of experimental data concerning the
Higgs sector, it is natural to suppose that there may be more than a
single Higgs boson waiting to be discovered at the weak scale.

The fundamental constraint on multi-Higgs doublet models comes from
flavor physics.  After diagonalizing the quark masses, the Yukawa
couplings of the neutral component of any extra Higgs boson will
generically lead to tree-level flavor-changing neutral current (FCNC)
processes, which are highly constrained by data.  There are only a few
options generally considered for avoiding these difficulties.  The
first is simply to assume that the Yukawa couplings of any new Higgs boson
to the standard model fermions are sufficiently small so as to be
safe.  Generically, for Higgs boson masses of order the weak scale, this
requires Yukawa couplings of order $10^{-4}$ or less.  The second
option is to demand that only a single Higgs boson couples to standard
model fermions of a given electric charge.  This leads to two commonly
considered scenarios, referred to as the Type I and II two Higgs
doublet models (2HDMs).  In the Type I 2HDM, it is assumed that an
additional Higgs does not couple to any of the standard model
fermions, while in the Type II model, a first Higgs couples only to
the up-type quarks, while a second couples only to the down-type
quarks, as in supersymmetry.  A third option often considered is that
of ``minimal flavor violation" \cite{MFV}.  In this scenario, it is assumed that
the full $U(3)^5$ flavor symmetry of the standard model is broken only
by the Yukawa couplings of a single Higgs boson responsible for
generating the fermion masses.  The Yukawas are assumed to come from
vacuum expectation values (vevs) of some set of fields transforming as
bifundamentals under the flavor group.  This results in all flavor
violation being of a size set by the ordinary standard model Cabbibo--Kobayashi--Maskawa (CKM)
matrix.

One feature which all of these scenarios have in common is that of
very small Yukawa couplings of the Higgs bosons to the first generation
quarks and leptons.  In the Type I and II 2HDMs this is required in
order to avoid giving large masses to the first generation fermions,
while in minimal flavor violation, this is required by virtue of the
smallness of the ordinary first-generation Yukawa couplings.  The smallness of these Yukawa couplings can
make it difficult, for example, to explain a recent $Wjj$ anomaly at
the Tevatron \cite{cdf} by using an extended Higgs sector.  In this paper we will
consider a simple alternative scenario which is able to avoid this
requirement, address the $Wjj$ anomaly, and simultaneously suggest a mechanism for a
straightforward solution to the strong CP problem.

Our setup assumes that a single Higgs boson $H$ dominates in providing the
masses for the standard model fermions.  Beyond this, we will make
two assumptions:
\begin{enumerate}
\item There is an $SU(3)_f$ flavor symmetry broken only by the $H$ Yukawa
  couplings.  The 3 right-handed up quarks, 3 right-handed down quarks, and 3
left-handed doublet quarks are all taken to transform in triplet representations
of the $SU(3)_f$ symmetry.  In particular, a second Higgs doublet,
  $\Phi=\left(\begin{array}{c} \phi^+ \\ \phi^0\end{array}\right)$ ,
  is assumed to have Yukawa couplings proportional to the identity
  matrix in a given canonical basis,\footnote{A canonical basis is defined to be the basis of quark fields where the flavor symmetry $SU(3)_f$ is manifest.} preserving the $SU(3)_f$.  Any further Higgs doublets beyond this should also have couplings of the same form.\footnote{Large diagonal Yukawa couplings for $H$ may
be suppressed in an appropriate UV completion.  See section 3 for an example.}
\item In the canonical basis, the $H$ Yukawa couplings are Hermitian matrices.
\end{enumerate}
%This second condition may easily arise, for example, if the H Yukawas
%come from the vacuum expectation values of fields transforming under
%the flavor $U(3)$ group in octet+singlet representations.

The reason that such a scenario can be associated with a solution to the strong
CP problem is straightforward; if CP is broken only
spontaneously, then the strong CP parameter, $\theta$, may be equal to zero
in the original canonical basis.  Diagonalizing the quark masses,
$\theta$ then {\it remains} equal to zero due to the Hermiticity
assumed for the Yukawa couplings.  Spontaneous breaking of CP could take place through a variety
of mechanisms already appearing in the literature, such as, for example, the Nelson/Barr mechanism \cite{NelsonCP, Barr}.  We must simply require that our "Hermitian Flavor Violation" (HFV) structure emerges in the effective theory at low energies.  This may be most easily accomplished if the $SU(3)_f$ flavor symmetry is respected by the sector of the UV theory responsible for CP violation.  It may also be possible to have the same fields simultaneously break both the flavor $SU(3)_f$ and CP symmetries together; a candidate for this type of theory will be discussed in section 3.

Due to the above assumed structure, the safety of the scenario from flavor-changing neutral currents
is simple to understand at the qualitative level.  After diagonalizing the quark masses,
the couplings of the neutral Higgs $\phi^0$ remain unchanged, while
the charged Higgs $\phi^+$ has flavor-changing interactions
proportional to the corresponding CKM elements.  In this way, the
structure of FCNC's is the same as in the standard model, with
analogous suppressions by the Glashow--Iliopoulos--Maiani (GIM) mechanism; we need only assume that
the $\Phi$ Yukawa is somewhat smaller than the gauge
coupling of the weak interaction, depending on the $\Phi$ mass.  The only difference here is that, in the presence of the $H$ Yukawa couplings, there is no symmetry fixing the universality of the
$\phi^0$ interactions. In the standard model, the $Z^0$ couplings
remain universal due to gauge invariance.  This will lead to some
small loop suppressed FCNCs, but not at a dangerous level.  We will
demonstrate the safety of the flavor structure in more detail in
section 2, as well as discuss the limits on the $\Phi$ Yukawa
couplings.\footnote{An analogous 2HDM flavor scenario was discussed in reference \cite{Argentina} but the motivation and underlying structure were different than what we consider here.}  Section 3 will contain a proposal for a possible UV completion of our scenario, demonstrating a mechanism for realizing the required hierarchical, Hermitian structure of the $H$ Yukawa couplings, as well as addressing the strong CP problem.

As noted above, the key phenomenological difference between this model
and more standard two Higgs doublet constructions is the presence of
allowed large couplings of $\Phi$ to the first generation fermions.
As a result, in section 4 we will discuss an explanation in this scenario for the excess in $Wjj$ events at CDF through resonant production of a heavy component of the new doublet.\footnote{For other thoughts in this direction, see \cite{Wang, Chen}} We will conclude in section 5.

\section{Flavor Constraints}

In the previous section, we presented a flavor structure which allows us to couple an additional Higgs boson to the standard model quarks, without having extremely suppressed couplings to the first generation.  In this section we examine the major constraints on this model and place limits on the couplings of the additional Higgs boson.  The Yukawa sector for this model is as follows:

\begin{equation}
-{\cal L}\supset  \tilde{H} {\bar Q_L}Y^U u_R +  H {\bar Q_L}Y^D d_R
    +\tilde{\Phi}  {\bar Q_L}G^U u_R + \Phi {\bar Q_L}G^D d_R,
\end{equation}
with $\tilde{H} = i \sigma_2 H^*$ and similarly for $\Phi$.
Here, the Yukawa couplings, $Y^U,Y^D,G^U,G^D$ are $3\times 3$ matrices.
We assume that $Y^U$ and $Y^D$ are Hermitian matrices and $G^U, G^D$ are proportional to the identity matrix ${\rm diag}(1,1,1)$ at some high-energy scale $\Lambda_{UV}$, with constants of proportionality $g^U$ and $g^D$ respectively.

One of the most important constraints on the model comes from the up and down quark masses.  Any term in the potential with an odd number of $\Phi$ fields (and a corresponding odd number of $H$ fields) will lead to a $\Phi$ vacuum expectation value and hence a contribution to the quark masses.   While such terms may be taken to be absent at tree level in appropriate UV completions (see the next section for an example), they will still be generated at loop level due to the $\Phi$ and $H$ Yukawa couplings.   Indeed,  the most important radiative corrections to the Higgs potential will come from top and bottom loops.  These lead to contributions
%Based on the UV completion presented below, we assume\footnote{This is enforced by a $Z_2$ symmetry which is then %broken.} $\langle \Phi\rangle\ne 0$ is only generated through radiative corrections.
\begin{equation}
{\cal L}_{mix}=-3\left(\frac{g^Dy_b+g^Uy_t}{8\pi^2}\right)\Lambda_{UV}^2\Phi^\dagger H +h.c.
\end{equation}
The quark mass contributions induced by this operator are
\begin{eqnarray}
\Delta m_d =-3 g^D\frac{g^Dm_b+g^Um_t}{8\pi^2}\frac{\Lambda_{UV}^2}{m_\phi^2}\\
\Delta m_u =-3 g^U\frac{g^Dm_b+g^Um_t}{8\pi^2}\frac{\Lambda_{UV}^2}{m_\phi^2},
\end{eqnarray}
where $m_\phi$ is the mass of the CP-even scalar component of $\Phi$. If we assume that these contributions are less than or equal to the values of the physical quark masses, then we may place the following approximate  upper limits on $g^U$ and $g^D$:
\begin{eqnarray}
 g^U \lesssim .007\left(\frac{m_\phi}{300 {\rm GeV}}\right)\left(\frac{{\rm TeV}}{\Lambda_{UV}}\right) \;\;\;\;  \\
  g^D\lesssim .06\left(\frac{m_\phi}{300 {\rm GeV}}\right)\left(\frac{{\rm TeV}}{\Lambda_{UV}}\right)\;\;\;\;
\end{eqnarray}
with the more severe constraint on $g^U$ due to $m_t/m_b\gg 1$. Although no tuning is needed when these constraints are satisfied, they can be relaxed if there is some degree of cancellation between the $\Phi$ and $H$ contributions to the quark masses.
% such that $Y'_{11}\langle H\rangle +G'\langle \Phi\rangle=m_d$.

We next examine flavor-changing neutral current constraints.  There are essentially two types of constraints we must consider.  The first type come from loop corrections to FCNC processes which are present even in the strict limit that the $\Phi$ couplings are proportional to the identity matrix.  By construction, these types of corrections have the same structure as in the standard model,  with $W^+$ or $H^+$ propagators replaced by $\phi^+$, and generally obtain similar GIM style suppressions.  In addition, we also have FCNC constraints coming from renormalization group (RG) running causing breaking of the perfect ${\rm diag}(1, 1, 1)$ forms of the $\Phi$ Yukawa couplings.  We consider these in turn.  Given the constraints already described above from the $\Phi$ vev requirement, the most important FCNC process we must consider is $K^0-\bar K^0$ mixing, and we will generally not discuss FCNC effects induced by the small coupling $g^U$.

%To the extent that the $\Phi$ couplings to the quarks remain diagonal after RG running, rotating to the mass eigenstate %basis does not affect the quark couplings to the neutral component of $\Phi$. However, it does lead to flavor mixed %couplings between the quarks for the charged Higgs boson. Fortunately, this new source of flavor violation is %proportional to the CKM matrix and a GIM like mechanism will suppress loop induced flavor violation from $\Phi^{\pm}$. %Without this GIM suppression, the flavor violation of this model would be to large. To verify that this GIM suppression is %sufficient, we examine the most severe constraints on HFV models below.

For $K^0-\bar K^0$ mixing, the experimental limits may be summarized as follows:
%  We state the constraints in terms of an effective cut off on the operator
%\begin{equation}
%\frac{\lambda}{\Lambda^2}(\bar d_L\gamma^\mu s_L)(\bar d_L\gamma_\mu s_L) + h.c. \label{EffOpCon}
%\end{equation}
%We can alter the above constraint to include $\bar s_L d_R$ by making the replacement $ \bar s_L d_R\to\bar %d_L\gamma^\mu s_L(m_{K^0}/m_s)$.
%The lower bounds on $\Lambda$ may then be written \cite{Blum:2009sk}:
%\begin{eqnarray}
%\Lambda\geq \sqrt{|\lambda|}~1000~{\rm TeV}
%\end{eqnarray}
%and
%\begin{eqnarray}
%\Lambda\geq \sqrt{|{\cal I}{\it m}(\lambda)|}~17000~{\rm TeV}.\label{IMbound}
%\end{eqnarray}
%For other operators also involving two $\bar{d}$ and two $s$ fields, the constraints will change in a calculable manner. 
 In the below analysis, we will find three types of induced operators,
\begin{eqnarray}
{\cal O}_{LL}= \bar s_R d_L \bar s_R d_L \;\;\;\;\;\;\;\;\;\; & {\cal O}_{RR}=\bar s_L d_R \bar s_L d_R \;\;\;\;\;\;\;\;\;\; & {\cal O}_{LR}= \bar s_R d_L \bar s_L d_R,
\end{eqnarray}
which all contribute to the $K^0-\bar K^0$ mixing.
%  Using the vacuum insertion approximation, it is possible to deduce the relative sizes of the contributions of these %operators to  $K^0-\bar K^0$ mixing  \cite{Branco:1999fs} in comparison with that of the operator in equation %\ref{EffOpCon}.  In this way, it follows that the required bounds on the scales $\Lambda$ in the above operators are %larger by factors of approximately
The current 95\%
confidence level bounds on these operators, expressed in terms of the required mass scale suppressing the real ($\Lambda_{\rm Re}$) and imaginary ($\Lambda_{\rm Im}$) parts, are \cite{constraints}
\begin{eqnarray}
{\cal O}_{LL},{\cal O}_{RR}: \;\;\;\;\;\;\;\; &\Lambda_{\rm Re}\geq 7.3\times10^3 {\rm TeV},  &\Lambda_{\rm Im}\geq 10^5 {\rm TeV}\label{LLRRbound}\\
{\cal O}_{LR}: \;\;\;\;\;\;\;\; & \Lambda_{\rm Re}\geq 1.7\times10^4 {\rm TeV},  &\Lambda_{\rm Im}\geq 2.4 \times 10^5 {\rm TeV} \label{LRbound}
\end{eqnarray}
%These enhancement factors from the vacuum insertion approximation have been shown to agree well with lattice results at energies of $\sim2$GeV for a %strange quark mass of $\sim 120$MeV \cite{Babich}, which is the value we will adopt here.\footnote{There may be some additional enhancements from% QCD effects, but we shall not take these into account here.  In any case, they would not cause a significant change to our results.}

\begin{figure}[t]
\centering
\includegraphics[width=0.47\columnwidth]{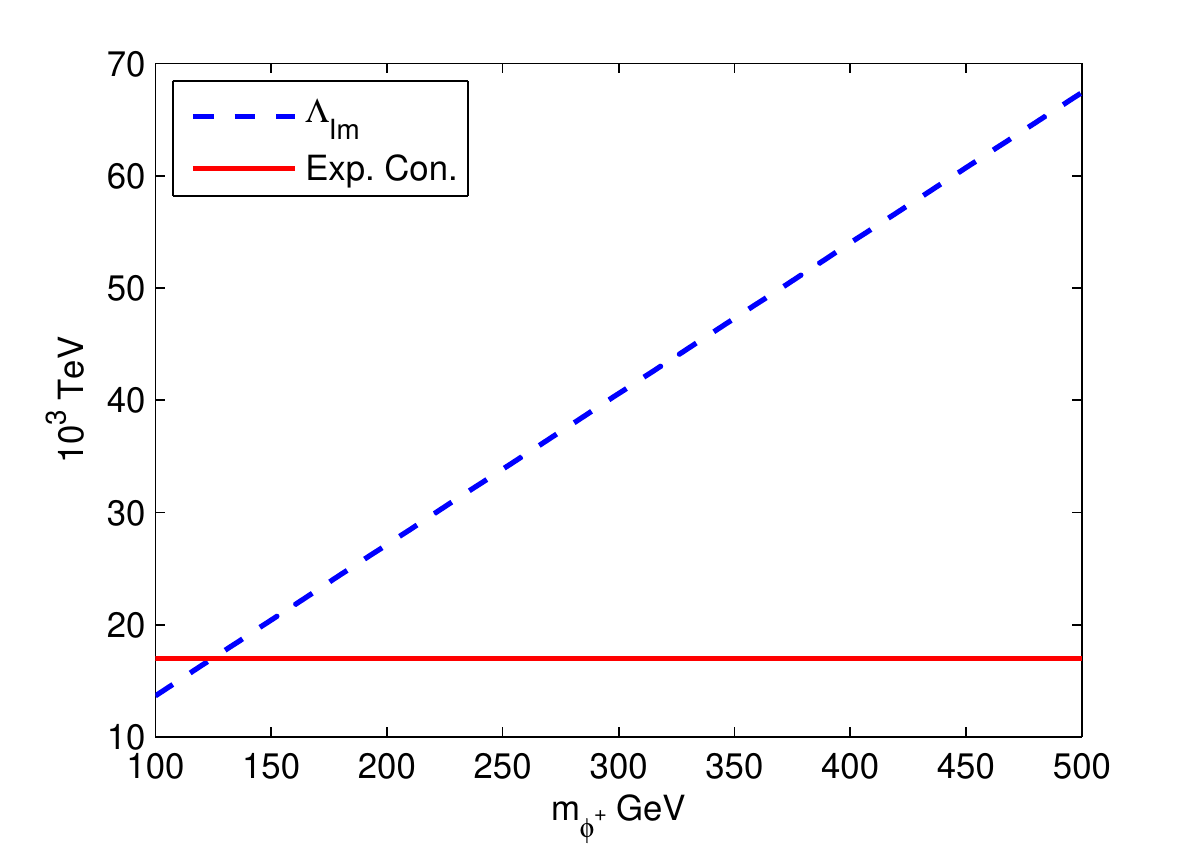}\includegraphics[width=0.47\columnwidth]{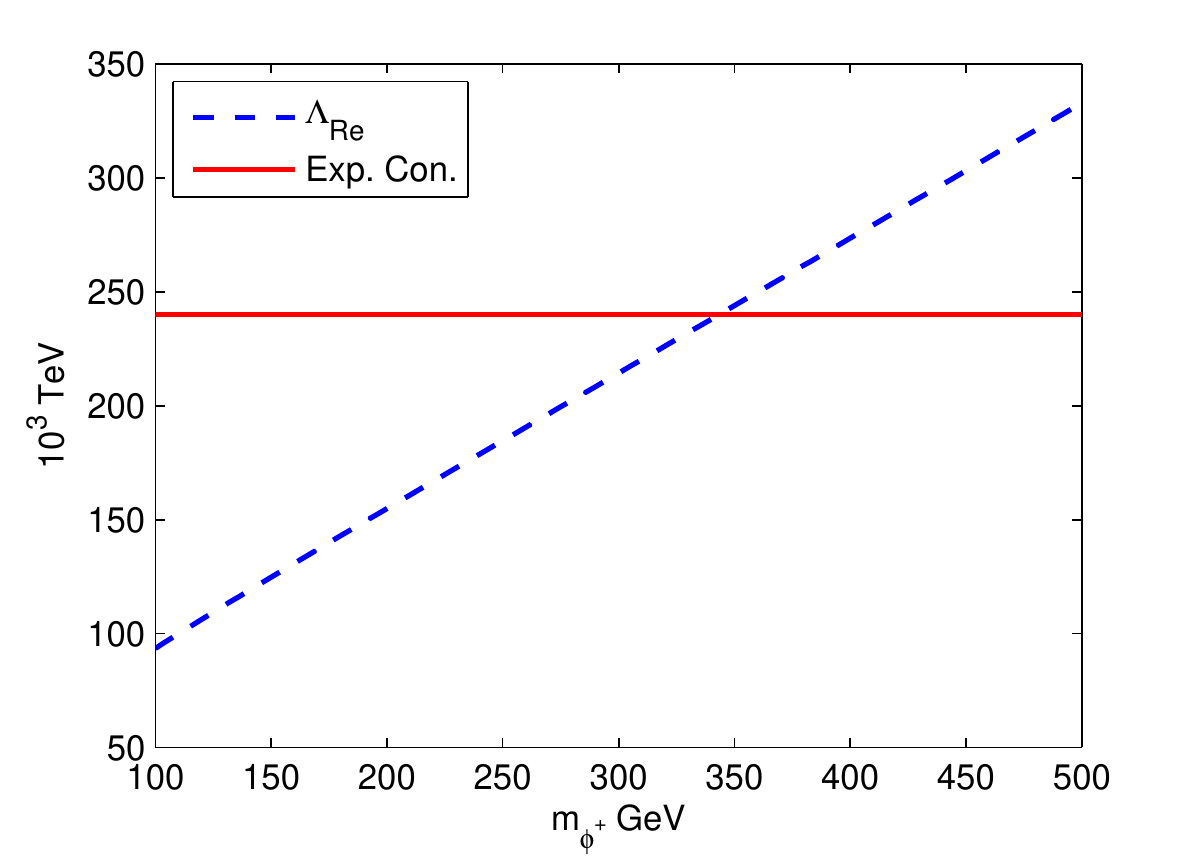}
\caption{We plot the mass scales appearing in the real (left) and imaginary (right) parts of the effective operator of type ${\cal O}_{LR}$ resulting from the $\phi^+$ box diagrams for a Yukawa coupling $g^D$ of 0.06.  The red lines show the constraints.  More generally, the values of $\Lambda_{\rm Re}$ and $\Lambda_{\rm Im}$ scale inversely with the coupling $g^D$. \label{KKMix} }
\end{figure}

We now consider the contribution to $K^0-\bar K^0$ mixing coming from replacing one or both of the $W^\pm$ lines with $\phi^\pm$ in the SM box diagrams. The expressions coming from these diagrams can be found in the Appendix.   The box diagrams contribute dominantly to $K^0-\bar K^0$ mixing through an induced operator of type ${\cal O}_{LR}$.  In   Fig. (\ref{KKMix}), we show the sizes of the mass scales in the real and imaginary parts of this operator in comparison with the experimental constraints, for a coupling $g^D$ of 0.06. 
%We plot the size of the induced  effective cut-off of these additional contributions in Fig. (\ref{KKMix}) for the operators
%\begin{eqnarray}
%\frac{1}{\Lambda_{Re}^2}\bar s_Ld_R\bar s_Rd_L+ \frac{i}{\Lambda_{Im}^2}\bar s_Ld_R\bar s_Rd_L
%\label{EffOp}
%\end{eqnarray}
Examining Fig. (\ref{KKMix}), we see that the constraint coming from $\Lambda_{\rm Im}$ is somewhat similar in severity to that coming from the size of the induced down quark mass.
% are for parameter regions which avoid the previously discussed constraint from the size of the down quark mass,  $K^0-\bar K^0$ mixing is safe.
%even for $g^D=.1$, which is a larger coupling than need to explain the $Wjj$ anomaly, and $m_{\phi}=150$ GeV, we %safely evaded the constraints on $K^0-\bar K^0$ mixing.

%These diagrams have already been computed
%in \cite{Argentina}, and the contribution with a single $\phi^\pm$ dominates. Because the  The effective operator which results from this loop calculation \cite{Argentina} takes the form
%
%\begin{eqnarray}
%\frac{g_2^2}{32\pi^2}\frac{g^{D^2}(\cos\theta_c\sin\theta_c)^2}{m_{\phi^+}^2}\frac{m_c^2}{m_W^2}
%\left(\left(\ln\left(\frac{m_W^2}{m_c^2}\right)-1\right)\bar{s}_L\gamma^\mu d_L\bar s_L\gamma_\mu d_L-\bar{s}_Ld_R\bar s_Ld_R\right)\\
%\nonumber = \left(\frac{g^D}{450~{\rm TeV}}\right)^2\bar{s}_L\gamma^\mu d_L\bar s_L\gamma_\mu d_L-\left(\frac{g^D}{1200~{\rm TeV}}\right)^2\bar{s}_Ld_R\bar s_Ld_R
%\end{eqnarray}
%where we have only kept leading order contributions, $g_2$  is the SU(2) gauge coupling, and $\theta_c$ is the Cabbibo angle. Although the second operator is naively a bit more suppressed, it will place the strongest constraint on $g^D$ due to the  $m_{K^0}/m_s$  enhancement.  In any case, however, it is apparent that this constraint is weaker than the one coming from the first generation quark masses.

As an aside, let us make a quick comment about the constraint coming from $b\to s\gamma$ decays. This constraint has been analyzed in \cite{Mahmoudi:2009zx} for a two Higgs doublet model where the additional Higgs boson couples to the SM fermions diagonally. This analysis can be applied to our scenario.  Taking $g^U$ small, we fall safely in the allowed parameter space for $g^D\lesssim 0.1$.\footnote{In \cite{Mahmoudi:2009zx} they state that for $\lambda_{tt}=0$ ($G^U_{33}$ in our notation, with $\lambda_{bb} = G^D_{33}$),
$b\to s\gamma$ decays are always safe.  This is an artifact of an approximation they make which allows them to neglect the $\lambda_{bb}^2$ contribution, which is not applicable to our case. However, the $b\to s\gamma$ contribution from the charged Higgs is invariant under an exchange $\lambda_{tt} \leftrightarrow \lambda_{bb}$ (with the dominant diagram undergoing a parity transformation).  In this way we may extract the limit for our case.}

We next consider the contribution to $K^0-\bar K^0$ mixing from RG running breaking the universality of the $\Phi$ couplings.  The dominant effect comes from the wave-function renormalization of the $Q_L$ fields, due to the large top Yukawa coupling to the $H$ doublet.
% Since only couplings with the SM Higgs boson, $H$, will break universality of $G^D$, we only need to consider the wave %function renormalization of the quark fields. The $d_R$ are only renormalized by the down quarks, and there wave %function renormalization will be suppressed by $y_b^2$. The $Q_L$, on the other hand, can have a wave function %renormalization that is proportional to the top Yukawa coupling.
Calculating the wave function renormalization of $Q_L$, we find
%The renormalization group equations for a theory with multiple Higgs bosons can be found in \cite{Cheng:1973nv}. The beta function for the $\Phi$ Yukawa coupling matrix $G^D$ is given by
%\begin{eqnarray}
%\beta_{Y_i}=\frac{1}{16\pi^2}
%\left(2Y_mY_iY_m+\frac{1}{2}\left(Y_mY_mY_i+Y_iY_mY_m\right)+2{\rm Tr}(Y_iY_m)Y_m\right)
%\end{eqnarray}
%where $Y_i$ is Hermitian Yukawa matrix for the Higgs boson $i$, repeated indices are summed over, and we have %neglected the gauge contributions to this equation. When applying this equation to the Yukawa coupling $G'$ of HFV we %get
\begin{equation}
\beta_{G^D}\supset \frac{G^D}{32\pi^2}\left(Y^UY^U\right),
\end{equation}
where we have neglected terms that are universal (since we only care about the breaking of universality here) and also smaller terms which are proportional to $Y^D$. We take the $\Phi$ Yukawa couplings to be universal at a UV scale $\Lambda_{UV}$, and then RG run down to the weak scale.   Without loss of generality, we are free to diagonalize the up quark mass matrix $Y^U$ at the UV scale, before performing the running. Since we already know that we must take $\Lambda_{UV}$ close to the TeV scale, we analyze the RG corrections using the leading-log approximation. The $G^D$ coupling at the EW scale is then

\begin{equation}
G^D(M_{EW})=G^D(\Lambda_{UV})+\frac{G^D}{32\pi^2}\left(Y^UY^U\right)\ln\left(\frac{\Lambda_{UV}}{M_{EW}}\right).
\end{equation}
After running the couplings to the weak scale, we then diagonalize $Y^D$ by redefining $d_{L_i}$ and $d_{R_j}$.  Although the corrections to $G^D$ are diagonal in the basis we did the RG running, they are not universal.  This non-universality gives family mixing when we rotate to the down quark mass eigenstates

\begin{equation}
V^\dagger G^D(M_{EW})V=G^D(\Lambda_{UV})+\frac{G^D}{32\pi^2}V^{\dagger}\left(Y^UY^U\right)V\ln\left(\frac{\Lambda_{UV}}{M_{EW}}\right).
\end{equation}
where $V$ is the CKM matrix.  We will now show that these radiative corrections are small enough to be less constraining than earlier bounds presented in this section. We neglect all of the present contributions to $K^0-\bar K^0$ mixing except those due to the top Yukawa coupling. The leading order contribution to the Lagrangian density resulting from scalar exchange is then
\begin{eqnarray}
%{\cal L}_{K^0-\bar K^0} \supset &&\!\!\!\!\!\!\!\!\!
\left(\frac{g^Dy_t^2V^*_{ts}V_{td}}{64\pi^2} \ln\left(\frac{\Lambda_{UV}}{M_{EW}}\right)\right)^2\left[\left(\frac{1}{m_\phi^2}-\frac{1}{m_A^2} \right)({\cal O}_{LL}+{\cal O}_{RR}) + 2\left(\frac{1}{m_\phi^2}+\frac{1}{m_A^2} \right){\cal O}_{LR} \right],
\end{eqnarray}
where  $A$ is the CP-odd scalar, with $m_A$ being its mass.  The coefficient of the ${\cal O}_{LL}+{\cal O}_{RR}$ operator then becomes
\begin{eqnarray}
\nonumber \left(\frac{g^D}{.1}\right)^2\left( \frac{(150 {\rm GeV})^2}{m_\phi^2}-\frac{(150 {\rm GeV})^2}{m_A^2}\right) &&\!\!\!\!\!\!\!\!\! \left(\left(\frac{\ln (\frac{\Lambda_{UV}}{M_{EW}})}{3\times 10^6{\rm TeV}}\right)^2-i\left(\frac{\ln(\frac{\Lambda_{UV}}{M_{EW}})}{3\times 10^6{\rm TeV}}\right)^2\right),
\end{eqnarray}
while that of the ${\cal O}_{LR}$ operator is
\begin{eqnarray}
\nonumber \left(\frac{g^D}{.1}\right)^2\left( \frac{(150 {\rm GeV})^2}{m_\phi^2}+\frac{(150 {\rm GeV})^2}{m_A^2}\right) &&\!\!\!\!\!\!\!\!\! \left(\left(\frac{\ln (\frac{\Lambda_{UV}}{M_{EW}})}{2\times 10^6{\rm TeV}}\right)^2+i\left(\frac{\ln(\frac{\Lambda_{UV}}{M_{EW}})}{2\times 10^6{\rm TeV}}\right)^2\right).
\end{eqnarray}
The strongest bounds on the above operators come from their imaginary parts, but by inspection of equations \ref{LLRRbound} and \ref{LRbound}, it is clear that they are quite a bit less constraining than the bounds coming from the scalar box diagram, or from the size of the induced down quark mass.
%In the above calculation we have used Wolfenstein parameterization of the CKM matrix. The SM is independent of the %parameterization of the CKM matrix, however,  HFV models are not. Above the scale where the horizontal symmetry is %broken, the theory is independent of the parameterization of the CKM matrix since only $\Phi$ couples to the SM quarks. %When the horizontal symmetry is broken, additional quark couplings are introduced. These additional couplings break the %parametrizational invariance of the theory and fix the form of the CKM matrix. The phase of $V_{23}V_{31}^\dagger$ is %determined by the breaking of the horizonal symmetry and could be suppressed relative to the Wolfenstein %parameterization. If the phase of this combination of CKM elements has a small phase, the strongest constraint would %come from the real part of the above operator. In which case, the constraint on $\Lambda$ from this operator is %negligible.

\section{A Possible UV Completion}

Here we present a UV completion which can realize our scenario in the IR. To get a Hermitian Flavor Violation model, the couplings of the SM Higgs doublet $H$ and the extra doublet $\Phi$ to the SM quarks must be very different;  $H$ obviously needs to have a very hierarchical Yukawa
matrix, while $\Phi$ must have an identity-like matrix.    In
addition, the $H$ Yukawa matrices are both required to be Hermitian.

The option we will consider here is to introduce a $Z_2$ symmetry to distinguish the two Higgs bosons, in addition to the $SU(3)_f$ flavor symmetry, with $Q_L$, $u_R$, and $d_R$  all taken as triplets under $SU(3)_f$ and even under $Z_2$.  We
assume $\Phi$ is even while $H$ is odd under $Z_2$.  In the symmetric
limit, $\Phi$ has Yukawa matrices with both the up- and down-sectors
proportional to the identity matrix, while $H$ does not have any
Yukawa couplings.  The Yukawa matrices $Y^U$ and $Y^D$ need to be
generated by exchange of heavy particles picking up
symmetry-breaking spurions.  We assume that the $SU(3)_f$ symmetry is
broken by vevs of triplet spurions $v_1$, $v_2$, and $v_3$, while the
$Z_2$ symmetry by a spurion $\sigma$.  CP is spontaneously broken in the
triplet VEVs, with the vev of $\sigma$ assumed real.  In order to communicate these symmetry breakings to the standard model sector, we introduce a set of Dirac fermions ${\cal U}$,
${\cal U}'_i$, ${\cal U}''_i$, ${\cal D}$, ${\cal D}'_i$, ${\cal D}''_i$, where  ${\cal U}$  and  ${\cal D}$ are flavor triplets, and where fields of type $i$, including $v_i$, are charged under separate $U(1)_i$ abelian symmetries.\footnote{We assume for simplicity that the flavor symmetries are gauged so that we don't have to worry about any light Goldstone modes, or possible Planck suppressed breaking effects.}   The ${\cal U}''_i$ and  ${\cal D}''_i$ fields are assumed even under the $Z_2$ symmetry, with all other new heavy quarks being odd.  We will label the left and right handed components of these Dirac fermions with $L$ and $R$ subscripts, as usual.  The full set of charges of the new fields are shown in Table 1.%\ref{tab:charges}.

\begin{table}[h]
  \centering
  \begin{tabular}{|c||c|c|c|c|}
    \hline
    Field & $SU(3)_f$ & $U(1)^3$ & ${\mathbb Z}_2$ & $SU(2)_W \times
    U(1)_Y $ \\ \hline
    ${\cal U}$ & 3 & 0 & $-$ & $1_{2/3}$ \\
    ${\cal U}'$ & 1 & 3 & $-$ & $1_{2/3}$ \\
    ${\cal U}''$ & 1 & 3 & $+$ & $1_{2/3}$ \\
    ${\cal D}$ & 3 & 0 & $-$ & $1_{-1/3}$ \\
    ${\cal D}'$ & 1 & 3 & $-$ & $1_{-1/3}$ \\
    ${\cal D}''$ & 1 & 3 & $+$ & $1_{-1/3}$ \\ \hline
    $H$ & 1 & 0 & $-$ & $2_{1/2}$ \\
    $\Phi$ & 1 & 0 & $+$ & $2_{1/2}$ \\
    $\sigma$ & 1 & 0 & $-$ & $1_0$ \\
    $v$ & 3 & 3 & $+$ & $1_0$ \\ \hline
  \end{tabular}
\label{tab:charges}
\caption{Charges of fields in the example UV completion.  Here a $U(1)^3$ charge of ``3" means that there are three such fields with separate $U(1)$ charges of the form $(1,0,0)$, $(0,1,0)$ and $(0,0,1)$. }
\end{table}

The following set of interactions are permitted by the symmetries, and will be used to generate the appropriate $H$ Yukawa structure (showing only the up sector for brevity):\footnote{Allowed heavy quark interactions with flipped chiralities may also be included and do not pose any difficulty.}
\begin{equation}
  {\cal O}_1 =  \tilde{H} \bar{Q}_L {\cal U}_R, \quad
  {\cal O}_2 = \bar{{\cal U}}_L v_i {\cal U}'_{R i}, \quad
 % {\cal O}_3 = \bar{{\cal U}}_- u \phi, \quad
  {\cal O}_3 = \sigma \bar{\cal U}'_{L i} {\cal U}''_{R i}, \quad
  {\cal O}_4 = \bar{\cal U}''_{L i} v^\dagger_i u_R.
\end{equation}
The resulting $H$ Yukawa couplings then have the form
\begin{equation}
  Y_u = \sum_i \frac{\langle v_i\rangle \langle \sigma\rangle \langle
    v_i\rangle^\dagger}{M_{{\cal U}} M_{{\cal
        U}'_i} M_{{\cal U}''_i}}\ , \qquad
  Y_d = \sum_i \frac{\langle v_i\rangle \langle \sigma\rangle \langle
    v_i\rangle^\dagger}{M_{{\cal D}} M_{{\cal
        D}'_i} M_{{\cal D}''_i}}\ .
\end{equation}
Note that there is no contribution that mixes up different $i$'s thanks to the $U(1)_i$ flavor symmetries.  This is crucial to ensure the Hermiticity of the Yukawa matrices.
Without loss of generality, we may make $SU(3)_f$ flavor rotations to put the $v$ vevs in form $v_1=(a,b,c)$, $v_2 =(d, e, 0)$, and $v_3 =(f, 0, 0)$.   Assuming an inverse hierarchy among the heavy fermion masses, we may then
obtain both Hermitian and hierarchical Yukawa matrices.  In this construction we may take the $\sigma$, $Z_2$ breaking vev to be of order TeV, along with one or more of the heavy quark masses, providing the effective $\Lambda_{UV}$ cutoff on the dangerous loop diagrams  discussed in section 2.   The schematic form
of the Yukawa matrices likely from this construction is
\begin{equation}
  Y^U \approx \left(
    \begin{array}{ccc}
      \lambda^8 & \lambda^8 & \lambda^8 \\
      \lambda^8 & \lambda^4 & \lambda^4 \\
      \lambda^8 & \lambda^4 & 1
    \end{array} \right), \qquad
  Y^D \approx y_b \left(
    \begin{array}{ccc}
      \lambda^3 & \lambda^3 & \lambda^3 \\
      \lambda^3 & \lambda^2 & \lambda^2 \\
      \lambda^3 & \lambda^2 & 1
    \end{array} \right),
\end{equation}
with $y_b$ being roughly the ratio of the bottom and top masses.

There are a few operators which are allowed by all of the symmetries of Table 1, but which are nevertheless dangerous to our construction.  These are
\begin{equation}
  \bar{Q}_L v^i i{\not\!\! D} v_i^\dagger Q_L, \qquad
  \sigma H^\dagger \Phi, \qquad
  \sigma \tilde{H} \bar{Q}_L u_R, \qquad
  \sigma \tilde{H}  \bar{Q}_L v_i v_i^\dagger v_j v_j^\dagger u_R.
\end{equation}
The first one leads to a non-universal Yukawa coupling to $\Phi$ and
hence flavor-changing neutral currents;  the second one induces a
vacuum expectation value for $\Phi$, too-large fermion masses and
hence fine-tuning; the third leads directly to too-large fermion masses;  the last one destroys the Hermiticity of the
Yukawa matrix if $i \neq j$.  Taking sufficiently suppressed coefficients for these operators is technically natural, so long as the $v_i$ vevs and masses are taken to be at least a few orders of magnitude larger than the TeV scale, with corresponding small coefficients for the   ${\cal O}_2$ and   ${\cal O}_4$ operators.  In general, the masses of the various fields, and coefficients of operators in this construction are somewhat flexible, and we will not discuss them in further detail.

It is a straightforward exercise to  check that this model leads to no strong CP parameter at tree level by considering the phase of the determinant of the full quark mass matrix.  At loop level, a highly suppressed contribution might arise after taking into account radiatively induced breaking of Hermiticity/universality, as well as the small induced $\Phi$ vev.  A full calculation of such loop corrections in this particular UV completion is however beyond the scope of the present work.

%  The realistic form
%of the Yukawa matrices likely from this construction is
%\begin{equation}
  %Y_u \approx \left(
   % \begin{array}{ccc}
    %  \lambda^8 & \lambda^8 & \lambda^8 \\
    %  \lambda^8 & \lambda^4 & \lambda^4 \\
    %  \lambda^8 & \lambda^4 & 1
    %\end{array} \right), \qquad
  %Y_u \approx y_b \left(
  %  \begin{array}{ccc}
   %   \lambda^3 & \lambda^3 & \lambda^3 \\
  %    \lambda^3 & \lambda^2 & \lambda^2 \\
  %    \lambda^3 & \lambda^2 & 1
 %   \end{array} \right).
%\end{equation}

%One possible concern is that the FN fermions may contribute to a
%deviation of the $\Phi$ Yukaws matrices from the identity matrix of
%the form $I + c Y_{u,d}$ with $c = O(1)$.  In this case, the FCNC
%processes may be somewhat larger than the earlier estimates, but are
%still under a good control at ``interesting'' levels.

%Another question is why the $\Phi$ doublet acquires very little VEV.
%We need $\langle \Phi \rangle \lesssim 10^{-5} \langle H \rangle$ in
%order not to overproduce the up-quark mass.  It requires a suppressed
%mass-mixing term $H^\dagger \Phi + \Phi^\dagger H$.  Assuming some
%mechanism to cancel quadratic divergences above $\Lambda \approx$~TeV,
%the mass mixing is generated at the level of $\frac{3}{16\pi^2} y_t
%h_u \Lambda^2$.  For $h_u \approx O(1)$, it requires a modest
%fine-tuning at the level of $10^{-2}-10^{-3}$.

\section{$Wjj$ events at the Tevatron}

The CDF collaboration recently reported on the production of $Wjj$
with an integrated luminosity of $4.3 {\rm ~fb}^{-1}$ \cite{cdf}.
Investigating the invariant mass distribution of the jet pair, they
found an excess of 253 events ($156\pm 42$ electrons, $97\pm 38$ muons)
in the $120-160$ GeV range, which is well fit by a Gaussian peak
centered at $144\pm 5$ GeV.  Additionally, it has been reported that analysis of an additional $3{\rm~fb}^{-1}$ sample collected by CDF shows this same feature,
%with a significance of $2.85\sigma$, giving a total significance close to $5\sigma$ for the combined $7.3{\rm~fb}^{-1}$ data set \cite{punzi}.
giving a total significance of $4.1\sigma$ for the combined $7.3{\rm~fb}^{-1}$ data set \cite{cdfnote}.

Hermitian Flavor violation provides a perfect setting for explaining the $Wjj$ anomaly
with a new $SU(2)$ doublet scalar.\footnote{Several explanations for the $Wjj$ anomaly have been presented in the literature, including other $SU(2)$ doublet scalars \cite{Wang, Babu, Dutta:2011kg}, a $Z$-prime \cite{Hooper, Cheung}, new colored states \cite{XPWang,Dobrescu,Carpenter:2011yj}, in supersymmetry \cite{Kilic, Sato}, technicolor \cite{Lane}, or string theory \cite{stringy}, and within the context of the Standard model \cite{He, Sullivan:2011hu,Plehn:2011nx}.}  Indeed, what seems to be required is a large coupling of the scalar to the first generation quarks.  However, as noted in the introduction, such couplings usually go hand in hand with
excessive flavor changing neutral currents.  Hermitian flavor violation addresses this problem.

Producing $Wjj$ events at the Tevatron through a new Higgs doublet can proceed via two primary mechanisms.  The first
is to simply have $t$-channel production of a $W$-boson along with the new scalar, followed by decay of the scalar to two jets.  This scenario, however, typically requires large couplings which run into difficulty with the constraints discussed in section 2 as well as collider constraints, and we will not discuss it further here. The second option is to split the masses of the charged and neutral Higgs components, and consider resonant production of the heavier state.  This state will then primarily decay into a $W$-boson plus the lighter state, with the lighter state then decaying to jets, as in Fig. (2)  (in this case there will also be a small additional $t$-channel contribution).
The CDF collaboration's background-subtracted invariant mass distribution of the $l\nu jj$ system,\footnote{The neutrino momentum may be reconstructed up to a two-fold ambiguity using the on-shell mass condition for the $W$.} $M_{l\nu jj}$, shows a peak in the $250\sim 300{\rm~GeV}$ range \cite{cdfnote}, which one would expect from a heavy resonance at around this mass range.
In general, we may consider some flexibility in the permitted scalar spectrum, pertaining to
the choice of which state is producing the final state jets, as well as whether or not the scalar and pseudo-scalar components are split from one another.  Hereafter we refer to the field(s) with mass $\sim 150{\rm~GeV}$ contributing to the excess as $\Phi_l$.

\begin{figure}[t]
\centering
\includegraphics[width=0.5\columnwidth]{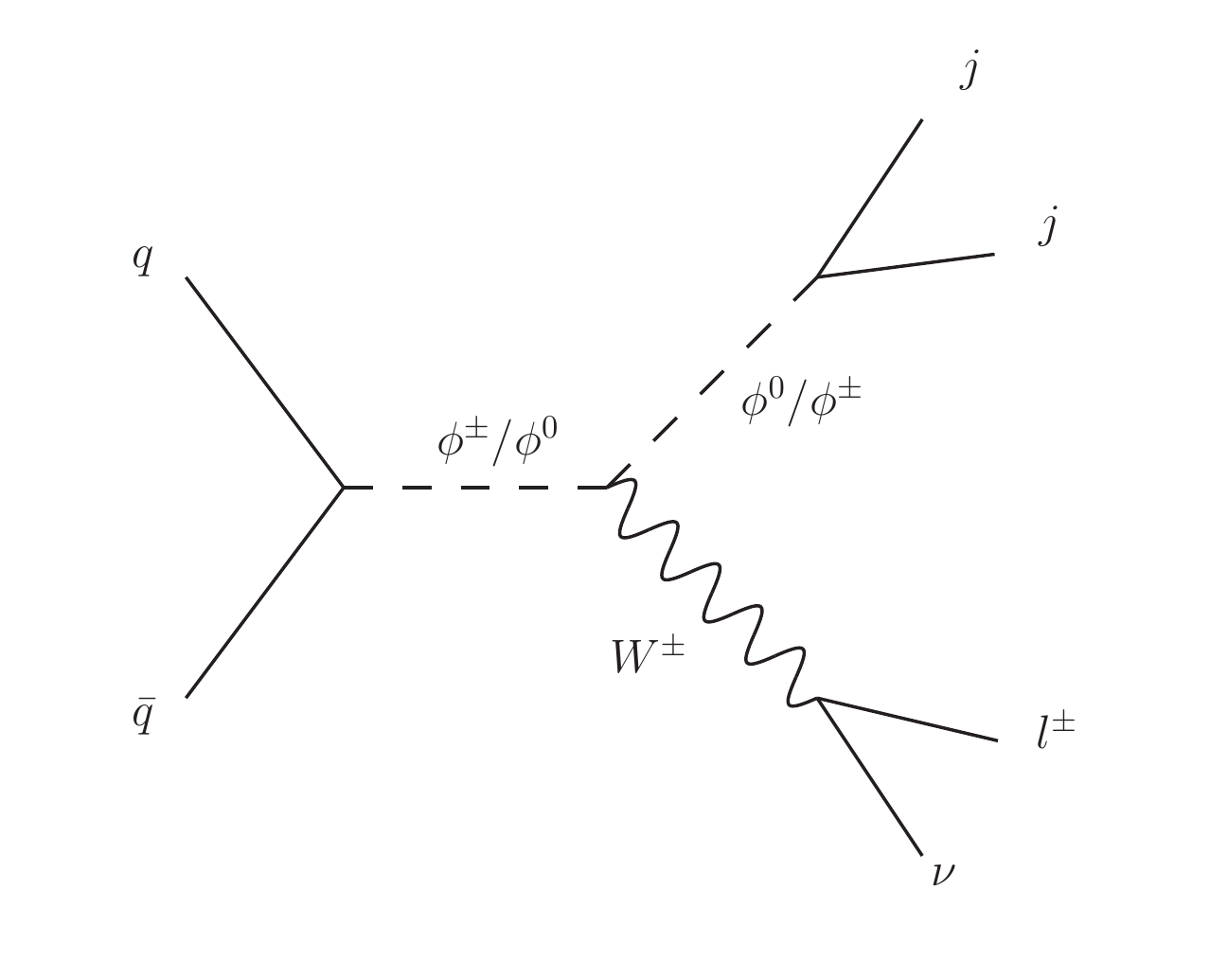}
\caption{Production of $W$ boson and two jets via resonant production. \label{fig:diagrams}}
\end{figure}

Since the $\Phi$ coupling to the $u_R$ quarks is relatively constrained by the requirement of not over-contributing to the up quark mass, as discussed in section 2, we will focus on the case where $\Phi$ couples dominantly to the $d_R$ quarks.\footnote{Though we focus on the case where $g^D$ is the dominant coupling, the results for coupling size in table 2 are essentially unchanged for the case of dominant $g^U$, with the exception of case 5, where the necessary initial partons required to produce a neutral resonance differ.  The $\sigma(Z\Phi_l)$, $\sigma(\gamma\Phi_l)$, and $\sigma(jj)$ entries in table 2, however, differ for the dominant $g^U$ case (due to different initial partons).}  Since the proton contains twice as many up quarks as down quarks, the best case for our model is to take $\phi^+$ to be the heavier, resonantly produced state.  At least one of the neutral components must then have a mass of $\sim 150$GeV in order to explain the CDF excess.
To get an idea of the size of the coupling needed to account for the $Wjj$ excess, we generate $p\bar{p}\to\Phi_l W^\pm\to l\nu jj$ events with Madgraph/MadEvent \cite{Alwall:2007st}, which are then showered with Pythia \cite{Sjostrand:2006za}, with detector simulation by PGS \cite{pgs} using CDF parameters.  We implement the cuts described in \cite{cdf}, and require a total of $\sim 250$ events to pass with a luminosity of $4.3{\rm~fb}^{-1}$.  We find good agreement between a $WW+WZ$ background created in this manner and the distribution in \cite{cdf}, suggesting that this provides a reasonable estimate. The results for several scenarios are shown in table 2.
For a variety of mass spectra, we see that the size of the required $\Phi$ Yukawa coupling is around $g^D \sim .06$.\footnote{This finding is consistent with the results of \cite{Wang}, who considered a phenomenologically similar model.}  As shown in section 2, a coupling of this size can evade all flavor constraints, as well as the constraint from the induced down quark mass, although the masses of $m_\phi$ and $m_\phi^+$ are preferred to be somewhat heavy, $\gsim 300 {\rm GeV}$.  These constraints might be relaxed somewhat after taking into account QCD uncertainties, or if we didn't require producing the central value of the CDF excess.  In particular, if we were to require producing only one standard deviation below the central value of the excess, then cases which previously required a coupling of .06 would instead require couplings of about .05.\footnote{The constraints from flavor could also be weakened if one were to adopt case ``5u" from Table 2.  In that case, $\Phi$ couples dominantly to the up sector, and fine tuning is required in order to keep the up quark mass small.  Such a fine tuning might be considered acceptable depending on one's perspective on the origin of the fermion mass hierarchy. }

\begin{table}
\begin{center}
\begin{tabular}{|c||c|c||c|c|c|c|}
\hline
{} & $m_{\phi}$, $m_{A}$, $m_{\phi^\pm}$ (GeV) &  $g^D$ &$\sigma(W^\pm \Phi_l)$ & $\sigma(Z \Phi_l)$ & $\sigma(\gamma \Phi_l)$ & UA2 $\sigma(jj)$\\ \hline
1 & 150 , 150 , 250 & 0.075 &4.1 pb & .032 pb & .008 pb& 2.0 pb\\ \hline
2 & 150 , 150 , 300 & 0.06 &1.7 pb & .020 pb & .005 pb& 1.3 pb\\ \hline
3 & 300 , 150 , 300 & 0.06 &1.6 pb& .430 pb & .003 pb& 0.6 pb\\ \hline
4 & 230 , 150 , 300 & 0.06 &1.7 pb& .016 pb & .003 pb& 0.6 pb\\ \hline
5d & 300 , 300 , 150 & 0.08 &1.6 pb& .028 pb & .016 pb& 5.2 pb\\ \hline
5u & 300 , 300 , 150 & (0.04) &1.5 pb& .008 pb & .004 pb& 1.3 pb\\ \hline
\end{tabular}
\label{tab:couplings}
\caption{Size of Yukawa couplings which explain the CDF $Wjj$ excess. The cross sections $\sigma(W^\pm \Phi_l)$, $\sigma(Z \Phi_l)$, and $\sigma(\gamma \Phi_l)$ are calculated at Tevatron energy, with no cuts apart from requiring $p_T>30$ for the photon. $\Phi_l$ refers to all fields with masses of $150{\rm~GeV}$.  $\sigma(jj)$ refers to the dijet cross section for the process $p\bar{p}\to\Phi_l\to jj$ at $\sqrt{s}=630{\rm~GeV}$ and should be compared with the limit of ${\mathcal O}(100{\rm~GeV})$ \cite{Alitti:1993pn}. The parentheses for model 5u indicate the value for the coupling $g^U$ rather than $g^D$.}
\end{center}
\end{table}

While table 2 presents only a few benchmark points, the behavior suggests that various mass spectra could in principle be able to explain the $Wjj$ excess.  There is some small variation in the required coupling with changes to the mass of the heavy resonance, as seen by comparison of scenarios 1 and 2.  This reflects both a larger branching ratio $BR(\phi^\pm\to W^\pm\phi^0)$ as well as a greater acceptance of events for the heavier resonance. 

If the scalar and pseudo-scalar masses are split, then the required coupling may change slightly, but not significantly, compared with the degenerate case.  As an example, consider taking the CP-even scalar component heavy, so that it is no longer within kinematic reach of the $\phi^+$ decays.  In that case, the size of the required $g^D$ coupling will remain essentially unchanged, at $\sim .06$, as seen by comparison of scenarios 2, 3, and 4 in table 2.  This follows because the width of the $\phi^+$ resonance, $\Gamma$, is cut in half.\footnote{For these scenarios, $\phi^\pm$ decays dominantly to $W^\pm\phi^0$, with $BR(\phi^\pm\to W^\pm\phi^0)\approx 96\%$ for scenario 2.}  Indeed, in the tree level production diagram, we obtain an increased resonant enhancement from a $1/\Gamma^2$ in the propagator, yielding a factor of 4.  There are half as many final states for the $\phi^+$ decay, yielding a suppression by a factor of 2.  Finally, due to the smaller width, there is half as much phase space volume for the initial quarks which can successfully hit the resonance. Taking into account the fact that the kinematics of the produced $Wjj$ events are unchanged from the degenerate case, and multiplying these factors together, we see that the overall event rate is essentially unchanged.

Aside from FCNC considerations, there are also direct collider constraints on two Higgs doublet models.
One might expect evidence of our additional Higgs sector in $\gamma jj$ and $Zjj$ events.  However, note that with resonant production, such events are quite suppressed, as shown in table 2, since the $\gamma$ and $Z$ cannot be produced in $\phi^+$ decays.  Scenario 3 has the largest $Zjj$ cross section because the CP-even scalar is heavy enough for the resonant process $d\bar{d}\to \phi\to A+Z$.  In contrast, the mass of $\phi$ in scenario 4 lies below the threshold for decay to $A+Z$, so it does not receive such an enhancement.

%In principle, the heavy resonance we consider could be seen within the $Wjj$ data, by reconstructing the neutrino using the $W$ mass and looking at the resulting $l\nu jj$ invariant mass.  In practice, however, because of the ambiguity of reconstructing the neutrino, the number of observed events would not be sufficient to expect a discernible peak.\cite{cdfthesis}

%Finally,
Additionally, a new scalar with
a coupling to first generation quarks could be produced as an $s$-channel
resonance and appear in dijet searches.  Because of the large QCD
backgrounds, Tevatron dijet bounds are only significant for resonances
heavier than those we consider here \cite{Aaltonen:2008dn}.  However,
the lower energy $p\bar{p}$ collisions ($\sqrt{s}=630 {\rm ~GeV}$)
observed by the UA2 collaboration provide an opportunity for
constraining ${\cal O}(100{\rm ~GeV})$ dijet resonances.  A search for
$W_R'$ resonances using a $10.9 {\rm~pb}^{-1}$ data sample places
constraints of ${\mathcal O}(100\rm~pb)$ for $\sigma\times BR(W'\to jj)$ at
the 90\% confidence level for a mass of $\sim 150{\rm
  ~GeV}$ \cite{Alitti:1993pn}.  Although we are considering a scalar
resonance, they provide a guideline for our extended Higgs
sector.  Our scenarios are very safe from this bound, as shown in table 2.

\section{Discussions and Conclusions}

In this paper, we presented a novel type of two-doublet Higgs model
that allows for new $O(.1)$ Yukawa couplings to the light generations while
naturally suppressing FCNCs via a GIM-like mechanism.   We also
discussed phenomenological consequences at colliders.  Thanks
to the allowed large couplings of the up- and down-quarks to the extra
doublet, the production of the doublet states can be significant.  In
particular, the bump in the $W+jj$ mass distribution reported by the
CDF collaboration may be explained straightforwardly in this setup, while remaining consistent with phenomenological constraints.
In addition, Hermiticity of the Yukawa couplings in our scenario suggests
a possible solution to the strong CP problem through spontaneous CP breaking.

\section*{Acknowledgements}

H.M. was supported in part by the U.S. DOE under Contract
DE-AC03-76SF00098, in part by the NSF under grant PHY-04-57315, and in
part by the Grant in-Aid for scientiﬁc research (C) 23540289 from
Japan Society for Promotion of Science (JSPS). The work of T.T.Y was supported by JSPS Grant-in-Aid for Scientific Research (A) (22244021). This work was also
supported by the World Premier International Center Initiative (WPI
Program), MEXT, Japan.

\appendix

\section{$K^0-\bar K^0$ Mixing Box Diagram}
Here we give a few details of the calculation of the $K^0-\bar K^0$ box diagrams with charged Higgs bosons. We first give the expression for the diagram with both a $\phi^\pm$ and $W^\pm_\mu$ which is
\begin{eqnarray}
{\cal L}\supset  \frac{(g_2g^D)^2}{16\pi^2}(\bar s_Ld_R\bar s_Rd_L) \frac{1}{m_\phi^2}
\sum_{i,j=1}^3 \lambda_i\lambda_j \left( F_A(x_W,x_i,x_j)+\frac{x_ix_j}{x_W}F_B(x_W,x_i,x_k)\right)\\
\nonumber
%+\frac{(g_2g^D)^2}{16\pi^2}(\bar s_L\sigma^{\mu\nu}d_R\bar s_R\sigma_{\mu\nu}d_L) \frac{1}{m_\phi^2}
%\sum_{i,j=1}^3 \lambda_i\lambda_j F_A(x_W,x_i,x_j)
\end{eqnarray}
where $\lambda_i=V_{is}^*V_{id}$, $g_2$ is the weak coupling constant, $x_i=m_i^2/m_{\phi}^2$, $m_i$ are the quarks masses, $x_W=m_W^2/m_\phi^2$, and
\begin{eqnarray}
F_A(x,y,z)=\int_0^1d^4z\frac{\delta(1-z_1-z_2-z_3-z_4)}{z_1+x z_2+y z_3 +z z_4}\\
F_B(x,y,z)=\frac{1}{2}\int_0^1d^4z\frac{\delta(1-z_1-z_2-z_3-z_4)}{(z_1+x z_2+y z_3 +z z_4)^2}
\end{eqnarray}

The box diagram with two charged Higgs bosons gives
\begin{equation}
{\cal L}\supset-\frac{(g^D)^4}{64\pi^2}(\bar s_R\gamma^\mu d_R\bar s_R\gamma_\mu d_R)\frac{1}{m_\phi^2}
\sum_{i,j=1}^3 \lambda_i\lambda_j  F_C(x_i,x_j)
\end{equation}
where
\begin{eqnarray}
F_C(x,y)=\int_0^1d^3z\frac{z_3\delta(1-z_1-z_2-z_3)}{z_1+x z_2+y z_3 }.
\end{eqnarray}
%Note that in the Vacuum Insertion Approximation, the operators with $\sigma^{\mu\nu}$ do not contribute to $K^0-\bar %K^0$ mixing.


\begin{thebibliography}{99}

%\cite{D'Ambrosio:2002ex}
\bibitem{MFV}
  G.~D'Ambrosio, G.~F.~Giudice, G.~Isidori and A.~Strumia,
  %``Minimal flavor violation: An Effective field theory approach,''
  Nucl.\ Phys.\  B {\bf 645}, 155 (2002)
  [arXiv:hep-ph/0207036].
  %%CITATION = NUPHA,B645,155;%%

\bibitem{cdf}
  T.~Aaltonen {\it et al.}  [CDF Collaboration],
  %``Invariant Mass Distribution of Jet Pairs Produced in Association with a $W$
  %boson in $p \bar{p}$ Collisions at $\sqrt{s}= 1.96$ TeV,''
  arXiv:1104.0699 [hep-ex].
  %%CITATION = ARXIV:1104.0699;%%

%\cite{Nelson:1983zb}
\bibitem{NelsonCP}
  A.~E.~Nelson,
  %``Naturally Weak CP Violation,''
  Phys.\ Lett.\  B {\bf 136}, 387 (1984).
  %%CITATION = PHLTA,B136,387;%%

%\cite{Barr:1984qx}
\bibitem{Barr}
  S.~M.~Barr,
  %``Solving the Strong CP Problem Without the Peccei-Quinn Symmetry,''
  Phys.\ Rev.\ Lett.\  {\bf 53}, 329 (1984).
  %%CITATION = PRLTA,53,329;%%

%\cite{Epele:1994jz}
\bibitem{Argentina}
  L.~Epele, H.~Fanchiotti, C.~Garcia Canal and D.~Gomez Dumm,
  %``FCNI suppression and CP violation in a two Higgs doublet model,''
  J.\ Phys.\ G {\bf 20}, 1159 (1994).
  %%CITATION = JPHGB,G20,1159;%%


  %If this reference is uncommented, its instance in the Wjj section should be commented
%\cite{Aaltonen:2011mk}




%\cite{Cao:2011yt}
\bibitem{Wang}
  Q.~H.~Cao, M.~Carena, S.~Gori, A.~Menon, P.~Schwaller, C.~E.~M.~Wagner and L.~T.~M.~Wang,
  %``W plus two jets from a quasi-inert Higgs doublet,''
  arXiv:1104.4776 [hep-ph].
  %%CITATION = ARXIV:1104.4776;%%

%\cite{Chen:2011wp}
\bibitem{Chen}
  C.~H.~Chen, C.~W.~Chiang, T.~Nomura and Y.~Fusheng,
  %``A light charged Higgs boson in two-Higgs doublet model for CDF $Wjj$
  %anomaly,''
  arXiv:1105.2870 [hep-ph].
  %%CITATION = ARXIV:1105.2870;%%

%The following references have been moved to Wjj section
%\cite{Wang:2011ta}
%\bibitem{XPWang}
%  X.~P.~Wang, Y.~K.~Wang, B.~Xiao, J.~Xu and S.~h.~Zhu,
%  %``New Color-Octet Vector Boson Revisit,''
%  arXiv:1104.1917 [hep-ph].
%  %%CITATION = ARXIV:1104.1917;%%
%%\cite{Buckley:2011vc}
%\bibitem{Hooper}
%  M.~R.~Buckley, D.~Hooper, J.~Kopp and E.~Neil,
%  %``Light Z' Bosons at the Tevatron,''
%  arXiv:1103.6035 [hep-ph].
%  %%CITATION = ARXIV:1103.6035;%%
%%\cite{Eichten:2011sh}
%\bibitem{Lane}
%  E.~J.~Eichten, K.~Lane and A.~Martin,
%  %``Technicolor at the Tevatron,''
%  arXiv:1104.0976 [hep-ph].
%  %%CITATION = ARXIV:1104.0976;%%
%%\cite{Kilic:2011sr}
%\bibitem{Kilic}
%  C.~Kilic and S.~Thomas,
%  %``Signatures of Resonant Super-Partner Production with Charged-Current
%  %Decays,''
%  arXiv:1104.1002 [hep-ph].
%  %%CITATION = ARXIV:1104.1002;%%
%%\cite{Cheung:2011zt}
%\bibitem{Cheung}
%  K.~Cheung and J.~Song,
%  %``Tevatron Wjj Anomaly and the baryonic $Z'$ solution,''
%  arXiv:1104.1375 [hep-ph].
%  %%CITATION = ARXIV:1104.1375;%%
%%\cite{He:2011ss}
%\bibitem{He}
%  X.~G.~He and B.~Q.~Ma,
%  %``The CDF dijet excess from intrinsic quarks,''
%  arXiv:1104.1894 [hep-ph].
%  %%CITATION = ARXIV:1104.1894;%%
%%\cite{Sato:2011ui}
%\bibitem{Sato}
%  R.~Sato, S.~Shirai and K.~Yonekura,
%  %``A Possible Interpretation of CDF Dijet Mass Anomaly and its Realization in
%  %Supersymmetry,''
%  arXiv:1104.2014 [hep-ph].
%  %%CITATION = ARXIV:1104.2014;%%
%%\cite{Anchordoqui:2011ag}
%\bibitem{stringy}
%  L.~A.~Anchordoqui, H.~Goldberg, X.~Huang, D.~Lust and T.~R.~Taylor,
%  %``Stringy origin of Tevatron Wjj anomaly,''
%  arXiv:1104.2302 [hep-ph].
%  %%CITATION = ARXIV:1104.2302;%%
%%\cite{Dobrescu:2011px}
%\bibitem{Dobrescu}
%  B.~A.~Dobrescu and G.~Z.~Krnjaic,
%  %``Weak-triplet, color-octet scalars and the CDF dijet excess,''
%  arXiv:1104.2893 [hep-ph].
%  %%CITATION = ARXIV:1104.2893;%%

%\cite{Nelson:2011us}
%\bibitem{Nelson}
%  A.~E.~Nelson, T.~Okui and T.~S.~Roy,
%  %``A unified, flavor symmetric explanation for the t-tbar asymmetry and Wjj
%  %excess at CDF,''
%  arXiv:1104.2030 [hep-ph].
%  %%CITATION = ARXIV:1104.2030;%%



%\cite{Aaltonen:2011kc}
%\bibitem{AFB}
 % T.~Aaltonen {\it et al.}  [CDF Collaboration],
  %``Evidence for a Mass Dependent Forward-Backward Asymmetry in Top Quark Pair
  %Production,''
 % arXiv:1101.0034 [hep-ex].
  %%CITATION = ARXIV:1101.0034;%%

%\cite{Peccei:1986pn}
%\bibitem{variant1}
 % R.~D.~Peccei, T.~T.~Wu and T.~Yanagida,
 % %``A VIABLE AXION MODEL,''
  %Phys.\ Lett.\  B {\bf 172}, 435 (1986).
 % %%CITATION = PHLTA,B172,435;%%


%%\cite{Bardeen:1986yb}
%\bibitem{variant2}
 % W.~A.~Bardeen, R.~D.~Peccei and T.~Yanagida,
  %%``CONSTRAINTS ON VARIANT AXION MODELS,''
  %Nucl.\ Phys.\  B {\bf 279}, 401 (1987).
 % %%CITATION = NUPHA,B279,401;%%

%\cite{Chen:2010su}
%\bibitem{variant3}
  %C.~R.~Chen, P.~H.~Frampton, F.~Takahashi and T.~T.~Yanagida,
 % %``Probing Variant Axion Models at LHC,''
 % JHEP {\bf 1006}, 059 (2010)
  %[arXiv:1005.1185 [hep-ph]].
 % %%CITATION = JHEPA,1006,059;%%

%%\cite{Chang:2008cw}
%\bibitem{hidden}
 % S.~Chang, R.~Dermisek, J.~F.~Gunion and N.~Weiner,
 % %``Nonstandard Higgs Boson Decays,''
  %Ann.\ Rev.\ Nucl.\ Part.\ Sci.\  {\bf 58}, 75 (2008)
  %[arXiv:0801.4554 [hep-ph]].
 % %%CITATION = ARNUA,58,75;%%

%%%%%%%%%%%%%%%%%%%%%%%%
% Flavor Section       %
%%%%%%%%%%%%%%%%%%%%%%%%

%\cite{Epele:1994jz}
%\bibitem{Epele:1994jz}
 % L.~Epele, H.~Fanchiotti, C.~Garcia Canal and D.~Gomez Dumm,
 % %``FCNI suppression and CP violation in a two Higgs doublet model,''
  %J.\ Phys.\ G {\bf 20}, 1159 (1994).
 % %%CITATION = JPHGB,G20,1159;%%

  %\cite{Blum:2009sk}
%\bibitem{Blum:2009sk}
  %K.~Blum, Y.~Grossman, Y.~Nir and G.~Perez,
  %%``Combining K0 - anti-K0 mixing and D0 - anti-D0 mixing to constrain the
  %%flavor structure of new physics,''
  %Phys.\ Rev.\ Lett.\  {\bf 102}, 211802 (2009)
  %[arXiv:0903.2118 [hep-ph]].
 % %%CITATION = PRLTA,102,211802;%%

%\cite{Branco:1999fs}
%\bibitem{Branco:1999fs}
 % G.~C.~Branco, L.~Lavoura and J.~P.~Silva,
 % %``CP Violation,''
  %Int.\ Ser.\ Monogr.\ Phys.\  {\bf 103}, 1 (1999).
 % %%CITATION = IMPHA,103,1;%%

%\cite{Babich:2006bh}
%\bibitem{Babich}
 % R.~Babich, N.~Garron, C.~Hoelbling, J.~Howard, L.~Lellouch and C.~Rebbi,
 % %``K0 - anti-0 mixing beyond the standard model and CP-violating electroweak
 % %penguins in quenched QCD with exact chiral symmetry,''
 % Phys.\ Rev.\  D {\bf 74}, 073009 (2006)
  %[arXiv:hep-lat/0605016].
 % %%CITATION = PHRVA,D74,073009;%%

%\cite{Bona:2007vi}
\bibitem{constraints}
  M.~Bona {\it et al.}  [UTfit Collaboration],
  %``Model-independent constraints on $\Delta$ F=2 operators and the scale of
  %new physics,''
  JHEP {\bf 0803}, 049 (2008)
  [arXiv:0707.0636 [hep-ph]].
  %%CITATION = JHEPA,0803,049;%%

%\cite{Mahmoudi:2009zx}
\bibitem{Mahmoudi:2009zx}
  F.~Mahmoudi and O.~Stal,
  %``Flavor constraints on the two-Higgs-doublet model with general Yukawa
  %couplings,''
  Phys.\ Rev.\  D {\bf 81}, 035016 (2010)
  [arXiv:0907.1791 [hep-ph]].
  %%CITATION = PHRVA,D81,035016;%%



 %\cite{Cheng:1973nv}
\bibitem{Cheng:1973nv}
  T.~P.~Cheng, E.~Eichten and L.~F.~Li,
  %``Higgs Phenomena in Asymptotically Free Gauge Theories,''
  Phys.\ Rev.\  D {\bf 9}, 2259 (1974).
  %%CITATION = PHRVA,D9,2259;%%




%%%%%%%%%%%%%%%%%%%%%%%%%%%%%%%%%%%%
% SECTION 4: Wjj Events at Tevatron%
%%%%%%%%%%%%%%%%%%%%%%%%%%%%%%%%%%%%

%\cite{Aaltonen:2011mk}
%\bibitem{cdf}
%  T.~Aaltonen {\it et al.}  [CDF Collaboration],
%  %``Invariant Mass Distribution of Jet Pairs Produced in Association with a $W$
%  %boson in $p \bar{p}$ Collisions at $\sqrt{s}= 1.96$ TeV,''
%  arXiv:1104.0699 [hep-ex].
%  %%CITATION = ARXIV:1104.0699;%%

%\cite{Buckley:2011vc}
%\bibitem{Buckley:2011vc}
%  M.~R.~Buckley, D.~Hooper, J.~Kopp, E.~Neil,
%  %``Light Z' Bosons at the Tevatron,''
%  [arXiv:1103.6035 [hep-ph]].


%\cite{Nelson:2011us}
%\bibitem{Nelson:2011us}
%  A.~E.~Nelson, T.~Okui and T.~S.~Roy,
%  %``A unified, flavor symmetric explanation for the t-tbar asymmetry and Wjj
%  %excess at CDF,''
%  arXiv:1104.2030 [hep-ph].
%  %%CITATION = ARXIV:1104.2030;%%

%\bibitem{punzi}
%  G.~Punzi, at 23rd Rencontres de Blois. [http://confs.obspm.fr/Blois2011/index.htm]
\bibitem{cdfnote}
  A. Annovia, P. Catastinib, V. Cavalierec, L. Ristorid, et. al., CDF note. [http://www-cdf.fnal.gov/physics/ewk/2011/wjj/7\_3.html].
  %Invariant Mass Distribution of Jet Pairs Produced in Association with a W boson in ppbar Collisions at $\sqrt{s}$ = 1.96 TeV

%\cite{Babu:2011yw}
\bibitem{Babu}
  K.~S.~Babu, M.~Frank, S.~K.~Rai,
  %``Top quark asymmetry and Wjj excess at CDF from gauged flavor symmetry,''
  [arXiv:1104.4782 [hep-ph]].

%\cite{Dutta:2011kg}
\bibitem{Dutta:2011kg}
  B.~Dutta, S.~Khalil, Y.~Mimura, Q.~Shafi,
  %``Dimuon CP Asymmetry in B Decays and Wjj Excess in Two Higgs Doublet Models,''
  [arXiv:1104.5209 [hep-ph]].

%\cite{Buckley:2011vc}
\bibitem{Hooper}
  M.~R.~Buckley, D.~Hooper, J.~Kopp and E.~Neil,
  %``Light Z' Bosons at the Tevatron,''
  arXiv:1103.6035 [hep-ph].
  %%CITATION = ARXIV:1103.6035;%%

%\cite{Cheung:2011zt}
\bibitem{Cheung}
  K.~Cheung and J.~Song,
  %``Tevatron Wjj Anomaly and the baryonic $Z'$ solution,''
  arXiv:1104.1375 [hep-ph].
  %%CITATION = ARXIV:1104.1375;%%


%\cite{Wang:2011ta}
%\bibitem{Wang}
\bibitem{XPWang}
  X.~P.~Wang, Y.~K.~Wang, B.~Xiao, J.~Xu and S.~h.~Zhu,
  %``New Color-Octet Vector Boson Revisit,''
  arXiv:1104.1917 [hep-ph].
  %%CITATION = ARXIV:1104.1917;%%

%\cite{Dobrescu:2011px}
\bibitem{Dobrescu}
  B.~A.~Dobrescu and G.~Z.~Krnjaic,
  %``Weak-triplet, color-octet scalars and the CDF dijet excess,''
  arXiv:1104.2893 [hep-ph].
  %%CITATION = ARXIV:1104.2893;%%

%\cite{Carpenter:2011yj}
\bibitem{Carpenter:2011yj}
  L.~M.~Carpenter, S.~Mantry,
  %``Color-Octet, Electroweak-Doublet Scalars and the CDF Dijet Anomaly,''
  [arXiv:1104.5528 [hep-ph]].

%\cite{Kilic:2011sr}
\bibitem{Kilic}
  C.~Kilic and S.~Thomas,
  %``Signatures of Resonant Super-Partner Production with Charged-Current
  %Decays,''
  arXiv:1104.1002 [hep-ph].
  %%CITATION = ARXIV:1104.1002;%%

%\cite{Sato:2011ui}
\bibitem{Sato}
  R.~Sato, S.~Shirai and K.~Yonekura,
  %``A Possible Interpretation of CDF Dijet Mass Anomaly and its Realization in
  %Supersymmetry,''
  arXiv:1104.2014 [hep-ph].
  %%CITATION = ARXIV:1104.2014;%%

%\cite{Eichten:2011sh}
\bibitem{Lane}
  E.~J.~Eichten, K.~Lane and A.~Martin,
  %``Technicolor at the Tevatron,''
  arXiv:1104.0976 [hep-ph].
  %%CITATION = ARXIV:1104.0976;%%

%\cite{Anchordoqui:2011ag}
\bibitem{stringy}
  L.~A.~Anchordoqui, H.~Goldberg, X.~Huang, D.~Lust and T.~R.~Taylor,
  %``Stringy origin of Tevatron Wjj anomaly,''
  arXiv:1104.2302 [hep-ph].
  %%CITATION = ARXIV:1104.2302;%%

%\cite{He:2011ss}
\bibitem{He}
  X.~G.~He and B.~Q.~Ma,
  %``The CDF dijet excess from intrinsic quarks,''
  arXiv:1104.1894 [hep-ph].
  %%CITATION = ARXIV:1104.1894;%%

%\cite{Sullivan:2011hu}
\bibitem{Sullivan:2011hu}
  Z.~Sullivan, A.~Menon,
  %``Standard model explanation of a CDF dijet excess in Wjj,''
  [arXiv:1104.3790 [hep-ph]].

%\cite{Plehn:2011nx}
\bibitem{Plehn:2011nx}
  T.~Plehn, M.~Takeuchi,
  %``W+Jets at CDF: Evidence for Top Quarks,''
  [arXiv:1104.4087 [hep-ph]].

%\cite{Alwall:2007st}
\bibitem{Alwall:2007st}
  J.~Alwall {\it et al.},
  %``MadGraph/MadEvent v4: The New Web Generation,''
  JHEP {\bf 0709}, 028 (2007)
  [arXiv:0706.2334 [hep-ph]].
  %%CITATION = JHEPA,0709,028;%%

%\cite{Sjostrand:2006za}
\bibitem{Sjostrand:2006za}
  T.~Sjostrand, S.~Mrenna, P.~Z.~Skands,
  %``PYTHIA 6.4 Physics and Manual,''
  JHEP {\bf 0605}, 026 (2006).
  [hep-ph/0603175].

\bibitem{pgs}
PGS 4, J. Conway et. al.

%\cite{Alitti:1993pn}
\bibitem{Alitti:1993pn}
  J.~Alitti {\it et al.} [ UA2 Collaboration ],
  %``A Search for new intermediate vector mesons and excited quarks decaying to two jets at the CERN $\bar{p} p$ collider,''
  Nucl.\ Phys.\  {\bf B400}, 3-24 (1993).

%%\cite{Cavaliere:2010zz}
%\bibitem{cdfthesis}
%%  V.~Cavaliere,
%  %``Measurement of ww + wz production cross section and study of the dijet mass spectrum in the lnu + jets final state at CDF,''
%	V.~Cavaliere (2010), Fermilab Ph.D Thesis 2010-51.
%  %%CITATION = FERMILAB-THESIS-2010-51;%%

%\cite{Aaltonen:2008dn}
\bibitem{Aaltonen:2008dn}
  T.~Aaltonen {\it et al.} [ CDF Collaboration ],
  %``Search for new particles decaying into dijets in proton-antiproton collisions at s**(1/2) = 1.96-TeV,''
  Phys.\ Rev.\  {\bf D79}, 112002 (2009).
  [arXiv:0812.4036 [hep-ex]].

\end{thebibliography}
\end{document}